\documentclass[aps,prl,nobibnotes,twocolumn,superscriptaddress,floatfix]{revtex4-2}

\usepackage{amsmath}
\usepackage{soul}
\usepackage{amssymb}
\usepackage{bm}
\usepackage{braket}
\usepackage{color}
\usepackage{graphicx}
\usepackage[colorlinks=true,pdfstartview=FitV,linkcolor=blue,citecolor=blue,urlcolor=blue]{hyperref}

\definecolor{BLUE}{rgb}{0,0,1}
\newcommand{\Ez}{\mathcal E_z}
\newcommand{\kb}{k_{\mathrm B}}
\newcommand{\sgn}{\mathrm{sgn}}

\begin{document}

\title{Unconventional Scaling of  Electric Hall Effect in Magnetic  Weyl Semimetals}

\author{Chaoxi Cui}
\affiliation{Centre for Quantum Physics, Key Laboratory of Advanced Optoelectronic Quantum Architecture and Measurement (MOE), School of Physics, Beijing Institute of Technology, Beijing 100081, China}
\affiliation{Beijing Key Lab of Nanophotonics \& Ultrafine Optoelectronic Systems, School of Physics, Beijing Institute of Technology, Beijing 100081, China}

\author{Yilin Han}
\affiliation{Centre for Quantum Physics, Key Laboratory of Advanced Optoelectronic Quantum Architecture and Measurement (MOE), School of Physics, Beijing Institute of Technology, Beijing 100081, China}
\affiliation{Beijing Key Lab of Nanophotonics \& Ultrafine Optoelectronic Systems, School of Physics, Beijing Institute of Technology, Beijing 100081, China}

\author{Run-Wu Zhang}
\affiliation{Centre for Quantum Physics, Key Laboratory of Advanced Optoelectronic Quantum Architecture and Measurement (MOE), School of Physics, Beijing Institute of Technology, Beijing 100081, China}
\affiliation{Beijing Key Lab of Nanophotonics \& Ultrafine Optoelectronic Systems, School of Physics, Beijing Institute of Technology, Beijing 100081, China}

\author{Zhi-Ming Yu}
\email{zhiming\_yu@bit.edu.cn}
\affiliation{Centre for Quantum Physics, Key Laboratory of Advanced Optoelectronic Quantum Architecture and Measurement (MOE), School of Physics, Beijing Institute of Technology, Beijing 100081, China}
\affiliation{Beijing Key Lab of Nanophotonics \& Ultrafine Optoelectronic Systems, School of Physics, Beijing Institute of Technology, Beijing 100081, China}

\author{Yugui Yao}
\affiliation{Centre for Quantum Physics, Key Laboratory of Advanced Optoelectronic Quantum Architecture and Measurement (MOE), School of Physics, Beijing Institute of Technology, Beijing 100081, China}
\affiliation{Beijing Key Lab of Nanophotonics \& Ultrafine Optoelectronic Systems, School of Physics, Beijing Institute of Technology, Beijing 100081, China}

\begin{abstract}
Electric Hall Effect (EHE), a unique phenomenon in two-dimensional (2D) magnetic systems, refers to the generation of Hall current by an out-of-plane electric
field $\Ez$. Here, we demonstrate that for  2D magnetic Weyl semimetals that host doubly degenerate nodal points,  the EHE features multiple unconventional scaling laws. At zero temperature, the EHE  exhibits a topological  $E_F^{-1}$ Fermi-energy scaling.
Remarkably,  the prefactor  of the scaling is determined by the global topological charge of the point without  any dependence on the local parameters of the system,   leading to  a universal and significant enhancement of Hall response in any species of Weyl points as the Fermi energy approaches the Weyl point. 
This significant response enables a weak electric field to be directly converted into a measurable Hall signal.
Surprisingly, this enhanced Hall response is not diminished by temperature, but evolves into  an unconventional logarithmically corrected scaling at finite temperature $\sigma_{xy}\propto\Ez\ln(1/|\Ez|)$ for  weak $\Ez$, still yielding a divergent electric-field susceptibility.
Thus, our work not only unveils   intriguing scaling laws resulting from the interaction between  magnetism and  topology, but also suggests a novel scaling-enhanced and temperature-robust mechanism that may enable weak electric-field sensing through a practical and all-electric route.
\end{abstract}

\maketitle

Universal scaling is a central concept of condensed matter physics, generally emerging near  the critical point of continuous phase transitions~\cite{Landau:1937obd,Fisher1967,Wilson1971,Hertz1976,Cardy1996,Sachdev2011}.  Classical  phase transitions occur at finite temperature, while  quantum phase transitions occur at  zero temperature, driven by physical parameter of system, such as pressure, doping and external fields~\cite{Hertz1976,HasanKane2010,Chandra2017,Mao2018,Cardy1996,Sachdev2011}.
Around a  critical point,  various kinds of  universal scaling laws emerge in thermodynamic and transport responses~\cite{Stanley1999,KouvelFisher1964,Ahlers1969,Wei1988}, such as susceptibilities, heat capacities, and conductivities. These scalings generally are insensitive to the  material details, and therefore provide a   powerful diagnostic for identifying the  transitions.
An important  feature of criticality is that certain susceptibilities diverge at the transition~\cite{Goldenfeld1992,ChaikinLubensky1995}, offering a natural route to amplify the corresponding response and, conversely, to detect the tuning field with high sensitivity.

Topological phase transitions form an important class of quantum phase transitions~\cite{Haldane1988,XiaoChangNiu2010, Bernevig2013,Shen2017,Chang2023,HasanKane2010,Armitage2018}, with criticality marked by a gap closing rather than a local order parameter~\cite{Bernevig2006,Konig2007,Xu2011, Yu2022Encyclopedia,Zhang2026FeTeO}.
In two dimensions (2D), Weyl points (WPs)~\cite{Zhang2023Encyclopedia,Young2015,Ahn2017}--referring to  doubly degenerate band crossings here--provide a minimal realization of such  topological critical point.
A WP generally appears as an unstable critical point separating two topologically distinct phases and requires fine tuning,  but additional crystalline symmetry can stabilize it~\cite{Zhang2023Encyclopedia}.
Besides, the WPs are topological nontrivial, hosting a nonzero  winding number~\cite{Novoselov2005,CastroNeto2009,Shen2017}.
Thus, the 2D Weyl semimetals  provide an ideal  platform for exploring topological critical scaling.

Meanwhile, modern electric field measurement faces a  challenge in miniaturization and integration~\cite{Horenstein1995,Noras2024,Liu2024EField,Zhang2024Rydberg}, because conventional electric field sensors generally  rely on auxiliary  transduction mechanisms, such as mechanically modulated charge induction and  electro-optic conversion.
In contrast, magnetic field sensing is considerably more mature~\cite{Lenz1990,Lenz2006,Ripka2010,Goel2020}.
The Hall sensor based on  ordinary Hall effect converts a magnetic field directly into a measurable transverse voltage without auxiliary transduction.
This immediately raises a fundamental question: can a similarly direct, all-electrical mechanism be devised for detecting weak electric field?

\begin{figure}[t]
	\includegraphics[width=\columnwidth]{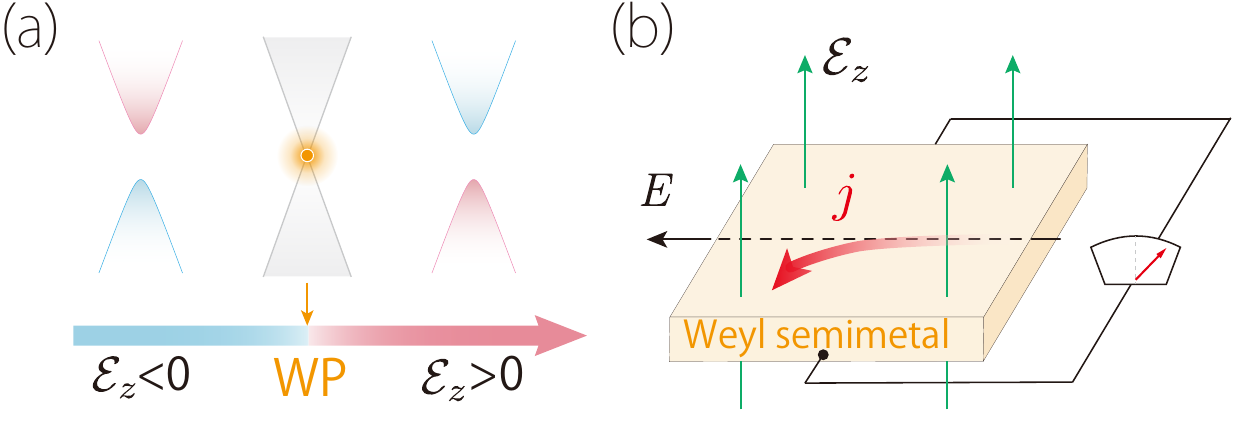}
	\caption{(a) $\Ez$-driven topological phase transition, where a 2D WP appearing at  $\Ez=0$ marks the gap-closing critical point between two gapped phases. 
		(b) Setup for electric-field sensing via the EHE in a 2D Weyl semimetal: an out-of-plane electric field $\Ez$ is directly read out via a Hall signal  $\boldsymbol j$ induced  by $\Ez$.}
	\label{Fig.1}
\end{figure}

The recently proposed electric Hall effect (EHE)~\cite{Cui2025EHE} provides a positive possibility. In the EHE, an out-of-plane electric field $\Ez$, rather than a magnetic field, generates a transverse Hall current in 2D magnetic systems. The EHE thus can convert the $\Ez$ directly into a Hall signal, making it a promising all-electric  mechanism for the measurement of weak electric field, provided that the response is sufficiently large.
A crucial  observation is that because the   stabilization of 2D WP requires additional crystalline symmetry, applying  $\Ez$ may break these  symmetries  and then gap the WP.
This process corresponds to a $\Ez$-driven topological phase transition, with the critical field being $\Ez=0$ [see Fig.~\ref{Fig.1}(a)].
Therefore, one can expect that around such critical points, i.e. $\Ez$-sensitive WPs,  the  susceptibilities of systems that are  coupled to $\Ez$, such as the EHE coefficient may feature divergent and  universal scaling.

In this work, we  establish the EHE scaling theory for generic $\Ez$-sensitive WPs at both zero and finite temperature.
At zero temperature ($T=0$), the EHE exhibits  a universal $E_F^{-1}$ Fermi-energy scaling with the prefactor determined  by the winding number of the WP (${\cal C}$), independent of the  model parameters.
This  leads to a universal and significant enhancement of Hall response in any species of $\Ez$-sensitive WPs.
More interestingly, at finite temperature, this enhancement, instead of being smeared out by thermal broadening, evolves into an unconventional logarithmically corrected scaling $\sigma_{xy}\propto\Ez\ln(1/|\Ez|)$ for weak $\Ez$, still yielding a divergent electric-field susceptibility.
Both scaling laws are universal and  guaranteed by topology, making the EHE near WPs a practical and all-electric route for sensing weak electric fields [see Fig.~\ref{Fig.1}(b)].

\textit{\textcolor{blue}{General formula for WP and EHE.}}
Consider a generic  two-band  model, for which the Hamiltonian can be expressed as
\begin{equation}\label{eq.two-band}
	H(\bm q)= {\bm d(\bm q)}  \cdot  {\bm \tau},
\end{equation}
where ${\bm q}=(q_x,q_y)$ is the momentum,  $\tau$'s are Pauli matrices, and ${d(\bm q)}$'s are  the coefficients determined by the details of the system.
Generally, the two-band model (\ref{eq.two-band}) is gapped. 
However, under certain critical conditions, such as additional crystalline symmetry or fine-tuning of external fields, the vector $\bm d$ can vanish at an isolated momentum point and the two-band model then becomes gapless, as illustrated in Fig.~\ref{Fig.1}(a).

This critical gapless point is exactly the WP, around which the most  generic   Hamiltonian up to the leading order can be written as~\cite{Zhang2023Encyclopedia}
\begin{equation}
	H_0(\bm k)= (c_1k_-^\nu+c_2k_+^\nu)\tau_+ + {\emph h. c.},
	\label{eq.HamWP}
\end{equation}
where  $\tau_{\pm}=(\tau_1 \pm i \tau_2)/2$,  $k_\pm=k_x\pm ik_y$ is the momentum measured from the WP, $c_1$ and $c_2$ are complex  parameters, and $\nu$ is an integer determining the order of the WP's dispersion. For a stabilized WP protected by crystalline symmetry,  $\nu$ can only be 1, 2, and 3, corresponding to linear, quadratic, and cubic WP, respectively.
The topological charge of the WP described by Eq. (\ref{eq.HamWP}) is the  winding number, given as  ${\cal C}=\nu\eta$ with $\eta=\sgn(|c_1|^2-|c_2|^2)$.


%


The field $\Ez$ may break the crystalline symmetry that protects the WP,  and then  gaps the WP by introducing a mass term
\begin{equation}
	H_{\mathcal E}=\Ez\hat{z}=m \tau_3,
	\label{eq.Hprime}
\end{equation}
where $m=\alpha\Ez$ is the effective mass with $\alpha$ a material-dependent coupling coefficient, generally determined by the height of the 2D system.
Moreover, when the global symmetries of the system do not forbid the EHE, the $\Ez$  will also induce finite Hall current.
We have  gone through the 528 magnetic layer groups (MLGs) and  screened out the ones that allow the existence of WPs and EHE~\cite{SM}. The results are listed in  Table~\ref{TabMLG}.
These candidates cover all three types of WPs with $\nu=1$, $2$, and $3$, respectively, and they can occur at both high-symmetry points and high-symmetry lines.

The EHE response is defined by
\begin{equation}
	j_a=\chi_{ab}\Ez E_b,
	\label{Eq1}
\end{equation}
where $\{a,b\}\in\{x,y\}$ with $a \ne b$. $\boldsymbol{E}$ and $\boldsymbol{j}$ denote the in-plane driving field and induced current in the 2D material, respectively.
$\chi_{xy}$ is the intrinsic EHE coefficient, given by~\cite{Cui2025EHE} (we set $h=e=1$)
\begin{align}
	\chi_{xy}=&\sum_n\int\frac{d^2k}{2\pi}
	\left(\partial_{\varepsilon_n} f \times
		P_{nn}\Omega_n
	+f \Lambda_n
	\right),
	\label{eq:chi}
\end{align}
where $f(\varepsilon_n-E_F)$ is the Fermi-Dirac distribution, $E_F$ is the Fermi energy, and $\varepsilon_n$ and $\ket{u_n}$ are the energy and Bloch state of the $n$-th band, respectively. $\Omega_n$ is the Berry curvature, defined as $\Omega_n=-2\,\mathrm{Im}\braket{\partial_{k_x}u_{n}|\partial_{k_y}u_{n}}$. $P_{nm}=\langle u_n|\hat z|u_m\rangle$ is the effective layer polarization.
$\Lambda_n$ denotes the Berry-curvature polarizability with respect to $\Ez$. For the two-band model described by Eq. (\ref{eq.HamWP}), it takes the form
\begin{align}
	\Lambda_n
	={}&2\operatorname{Im}
	\frac{1}{(\delta \varepsilon_{n \bar{n}})^3}
	\Big[
	2\delta P_{n\bar{n}} v_x^{n\bar n}v_y^{\bar n n}
	\nonumber\\
	&\quad+
	P_{\bar n n}\delta V_y^{ n\bar{n}} v_x^{n\bar n}+
	P_{n\bar n}\delta V_x^{ n\bar{n}} v_y^{\bar n n}
	\Big],
	\label{eq:Lambda}
\end{align}
where $n$ and $\bar n$ are the band index with $n\neq \bar n$, $v_a^{nm}=\langle u_n|\partial_{k_a} H_0|u_m\rangle$. We have defined $\delta \varepsilon_{n \bar{n}}=\varepsilon_n-\varepsilon_{\bar n}$, $\delta P_{ n \bar{n}}=P_{nn}-P_{\bar n\bar n}$, and $\delta V_a^{n\bar{n} }=v_a^{nn}-v_a^{\bar n\bar n}$. 
For computational convenience, we can take the $n$-th ($\bar n$-th) band to be occupied (empty), because when  both bands are occupied, we always have  $\Lambda_n+\Lambda_{\bar n}=0$.

\textit{\textcolor{blue}{Scaling at zero temperature.}}
Before performing detailed calculations, we first demonstrate  that a simple scaling analysis can directly show that $\chi_{xy}$ of the WPs described by Eq. (\ref{eq.HamWP}) follows a universal $E_F^{-1}$ scaling.

Consider the following scaling transformation in momentum and energy~\cite{Cao2024CPME,Ahn2020BPVE,Li2023PHE}:
\begin{equation}
	\bm k\rightarrow\lambda^{1/\nu}\bm k,
	\qquad
	E_F\rightarrow\lambda E_F,
	\label{eq:scaling_transform}
\end{equation}
with $\lambda>0$. Under this transformation, the Hamiltonian and band energies satisfy
$H_0(\lambda^{1/\nu}\bm{k})=\lambda H_0(\bm{k})$ and
$\varepsilon_n(\lambda^{1/\nu}\bm{k})=\lambda\varepsilon_n(\bm{k})$.
The eigenstates depend only on the direction of $\bm{k}$, and hence remain invariant under the rescaling, i.e. $\ket{u_n(\lambda^{1/\nu}\bm{k})}=\ket{u_n(\bm{k})}$.
It follows that
$P_{nm}(\lambda^{1/\nu}\bm{k})=P_{nm}(\bm{k})$.
By contrast, the momentum derivative scales as
$\partial_{k_a}\rightarrow\lambda^{-1/\nu}\partial_{k_a}$.
Therefore, the velocity matrix elements and Berry curvature satisfy
$v_a^{nm}(\lambda^{1/\nu}\bm{k})=\lambda^{1-1/\nu}v_a^{nm}(\bm{k})$
and
$\Omega_n(\lambda^{1/\nu}\bm{k})=\lambda^{-2/\nu}\Omega_n(\bm{k})$.
Combining these scaling relations in Eq.~(\ref{eq:Lambda}) gives
$\Lambda_n(\lambda^{1/\nu}\bm{k})=\lambda^{-1-2/\nu}\Lambda_n(\bm{k})$.

At zero temperature, $f=\Theta(E_F-\varepsilon_n)$, with $\Theta$ the step function, is invariant under Eq.~(\ref{eq:scaling_transform}), while $\partial f/\partial\varepsilon_n$ scales as $\lambda^{-1}$. Since $d^2k\rightarrow\lambda^{2/\nu}d^2k$, the integral of the Fermi-surface term $\partial_\varepsilon f\,P_{nn}\Omega_n$ and the Fermi-sea term $f\Lambda_n$ in Eq.~(\ref{eq:chi})  acquires the  factor $\lambda^{-1}$. Hence, the zero-temperature EHE coefficient $\chi_{xy}^{(0)}$ features  a  simple scaling relation:
\begin{equation}
	\chi_{xy}^{(0)}(\lambda E_F)
	=\lambda^{-1}\chi_{xy}^{(0)}(E_F),
	\label{eq:chi_scaling}
\end{equation}
which  directly  gives  that $\chi_{xy}^{(0)}$ scales as $E_F^{-1}$. 

To further study the nontrivial properties of the scaling, we  establish the full expression of  $\chi_{xy}^{(0)}$  via Eq.~(\ref{eq:chi}).
For the $\Ez$-sensitive WPs,  the  effective layer polarization  can be expressed  as $P_{nm}=\langle u_n|\hat z|u_m\rangle=\alpha \langle u_n|\tau_3|u_m\rangle$ consistent with the expression of Eq. (\ref{eq.Hprime}).
Then, via  straightforward calculations~\cite{SM}, we establish the EHE of the WP described by Ham. (\ref{eq.HamWP}) at $T=0$, expressed as
\begin{equation}\label{eq:chi_zeroT}
	\chi_{xy}^{(0)}(E_F) = \frac{\alpha}{2} \times	\frac{{\cal C}}{|E_F|}.
\end{equation}

\begin{figure}[t]
	\includegraphics[width=\columnwidth]{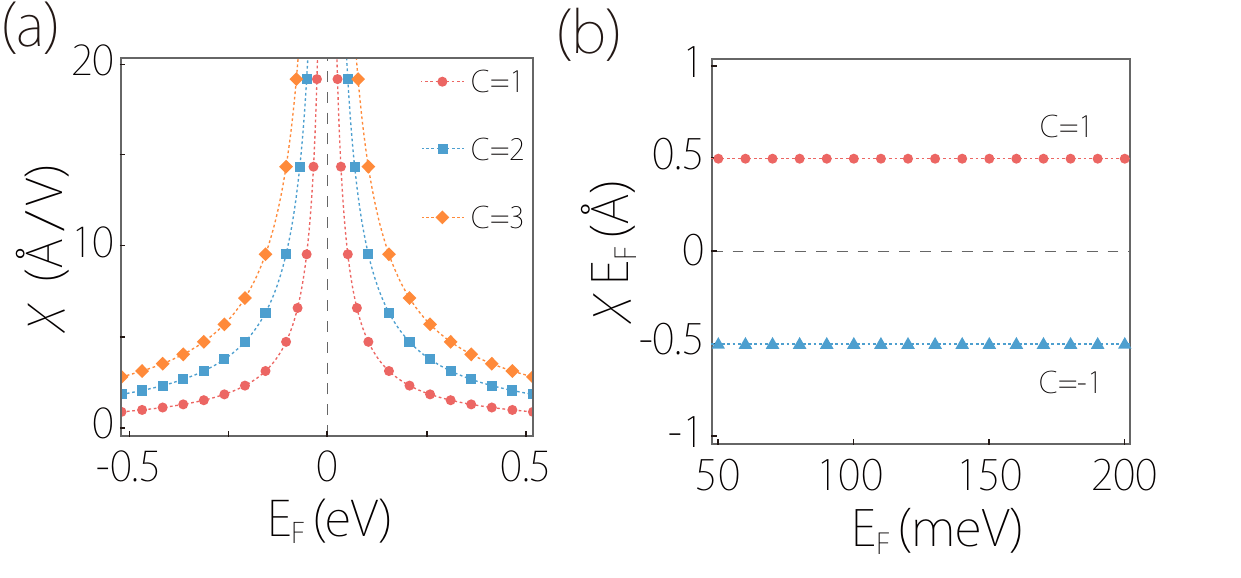}
	\caption{ EHE response at zero temperature. 
		(a) $\chi_{xy}^{(0)}$  for WPs with topological charges ${\cal C}=1,2$, and $3$.  
		(b) $\chi_{xy}^{(0)}E_F$ for WPs with ${\cal C}=\pm1$. 
$\chi_{xy}^{(0)}$, which scales with $E_F^{-1}$, is independent of the details of the WPs, but  depends on  the global topological charge ${\cal C}$.
		Here we set $\alpha=1~\text{\AA}$.}
	\label{Fig.2}
\end{figure}

This remarkable  result indicates several points. First, the underlying physics is as expected:  $\chi_{xy}^{(0)}$ is the susceptibility of Hall conductivity with respect to $\Ez$, and then exhibits a universal expression around the critical point of a  $\Ez$-driven topological phase  transition, which here are the  WPs.
It does not depend on  the details of the WPs, such as $c_1$ and $c_2$, but is fully determined by the global property, i.e. topological charge ${\cal C}$ of the WPs, as shown in Fig.~\ref{Fig.2}(b).
Second,  the   EHE and the Hall signal can be significantly enhanced when  the Fermi energy approaches the nodal point.
Third, the EHE is more significant for the WPs with quadratic and cubic dispersion [see Fig.~\ref{Fig.2}(a)].

\textit{\textcolor{blue}{Scaling at finite temperature.}}
Since  temperature can fundamentally reshape the responses near critical points, we further  investigate the evolution of topological EHE scaling near a 2D WP at finite temperature.

Since the expression of  EHE at zero temperature is obtained,   one may attempt to obtain its finite-temperature counterpart via the spectral convolution~\cite{Maldague1978,SmrckaStreda1977}
\begin{equation}
	\chi_{xy}^{(T)}(E_F)
	=	\int_{-\infty}^{\infty}d\varepsilon \chi_{xy}^{(0)}(\varepsilon)  {\cal B}(\varepsilon-E_F,T),
	\label{eq:chiT_divergent}
\end{equation}
where ${\cal B}(\varepsilon-E_F,T)=-\partial_\varepsilon f(\varepsilon-E_F,T)$ is the thermal broadening kernel, which  is concentrated within an energy window of order $\kb T$ around $E_F$, and decays exponentially for large $|\varepsilon-E_F|$. 
However, the integration in Eq.~(\ref{eq:chiT_divergent}) inevitably samples $\varepsilon=0$, where $\chi_{xy}^{(0)}(\varepsilon)$ is divergent.

Fortunately, the $\Ez$-induced zero-temperature Hall conductivity   $\sigma_{xy}^{(0)}$ can be well defined regardless of the value of Fermi energy, as it is a physical observable.
For the WP described by Ham. (\ref{eq.HamWP}), $\sigma_{xy}^{(0)}$ takes the form of  $\sigma_{xy}^{(0)}=\chi_{xy}^{(0)} \Ez$ for $|\alpha\Ez|<|E_F|$, but   $\sigma_{xy}^{(0)}=\frac{{\cal C}}{2}  \times \sgn(\alpha \Ez)$ for $|\alpha\Ez|>|E_F|$~\cite{SM}.
Since  $\chi_{xy}^{(0)}=\partial_{\Ez}\sigma_{xy}^{(0)}$ is well defined in the weak-field limit, the condition $|\alpha\Ez| < |E_F|$ is always satisfied for any finite $E_F$, and  Eq.~(\ref{eq:chi_zeroT}) is valid except $E_F=0$.

Applying the spectral convolution to $\sigma_{xy}^{(0)}$, we  obtain the finite-temperature Hall conductivity
\begin{align}
	\sigma_{xy}^{(T)}
	={}&
    \frac{{\cal C}}{2} \times	\sgn(\alpha\Ez)
	\int_0^{|\alpha\Ez|}
	d\varepsilon\,
	\mathcal W(\varepsilon,E_F,T)
	\nonumber\\
	&+
	\frac{{\cal C}}{2} \times \alpha\Ez
	\int_{|\alpha\Ez|}^{\infty}
	\frac{d\varepsilon}{\varepsilon}
	\mathcal W(\varepsilon,E_F,T),
	\label{eq:sigma_full}
\end{align}
with $\mathcal W(\varepsilon,E_F,T)={\cal B}(\varepsilon+E_F,T)+{\cal B}(\varepsilon-E_F,T)$. 
In the weak-field limit of $|\alpha\Ez|\ll k_BT$, the   first term of Eq.~(\ref{eq:sigma_full}) contributes only a regular linear  term $\varpropto \Ez$, as $\mathcal W(\varepsilon,E_F,T)$ is an analytic  function without singularity. 
For the second term in  Eq.~(\ref{eq:sigma_full}), we can rewrite it as 
\begin{align}\label{eq:sigma_full2}
	\frac{\alpha\Ez{\cal C}}{2}	\int_{|\alpha\Ez|}^{\infty} \frac{d\varepsilon}{\varepsilon}	\mathcal W= \frac{\alpha\Ez{\cal C}}{2}	\left(\int_{|\alpha\Ez|}^{\varepsilon^{\prime}} +\int_{\varepsilon^{\prime}}^{\infty} \right)\frac{d\varepsilon}{\varepsilon}	\mathcal W,
\end{align}
 where  $\varepsilon^{\prime}\gg|\alpha\Ez|$ is  an intermediate parameter.  
Since  ${\cal W}(\varepsilon,E_F,T)$  decays exponentially for large $\varepsilon$, the latter term in   Eq.~(\ref{eq:sigma_full2}) also contributes a regular term $\varpropto |\Ez|$. In contrast, because $1/\varepsilon$ is significant only around $|\alpha\Ez|$ ($\simeq 0$), the former term in  Eq.~(\ref{eq:sigma_full2}) generally can be approximately expressed as~\cite{Hinch1991,Mahan2000}  
\begin{align}
&\frac{\alpha\Ez{\cal C}}{2} \int_{|\alpha\Ez|}^{\varepsilon^{\prime}}  \frac{d\varepsilon}{\varepsilon} \mathcal W \approx	\frac{\alpha\Ez{\cal C}}{2}	 \mathcal W(0,E_F,T) \int_{|\alpha\Ez|}^{\varepsilon^{\prime}}  \frac{d\varepsilon}{\varepsilon}	\nonumber\\
&=\frac{\alpha\Ez{\cal C}}{2}	 \mathcal W(0,E_F,T) \ln (1/|\Ez|)+ O(\Ez),
\end{align}
with $O(\Ez)$ denoting the term that can be neglected compared with  the first term $\varpropto \ln (1/|\Ez|) \Ez$ in the limit of weak  $\Ez$. 

Consequently, to the leading order of $\Ez$,  we obtain a simple formula for the finite-temperature Hall conductivity, 
\begin{align}\label{eq:sigma_log}
	\sigma_{xy}^{(T)}	=\frac{\alpha{\cal C}}{2}	 \mathcal W(0,E_F,T) \ln (1/|\Ez|)\Ez+ O(\Ez),
\end{align}
which  is the second important result of this work.

\begin{figure}[t]
	\includegraphics[width=\columnwidth]{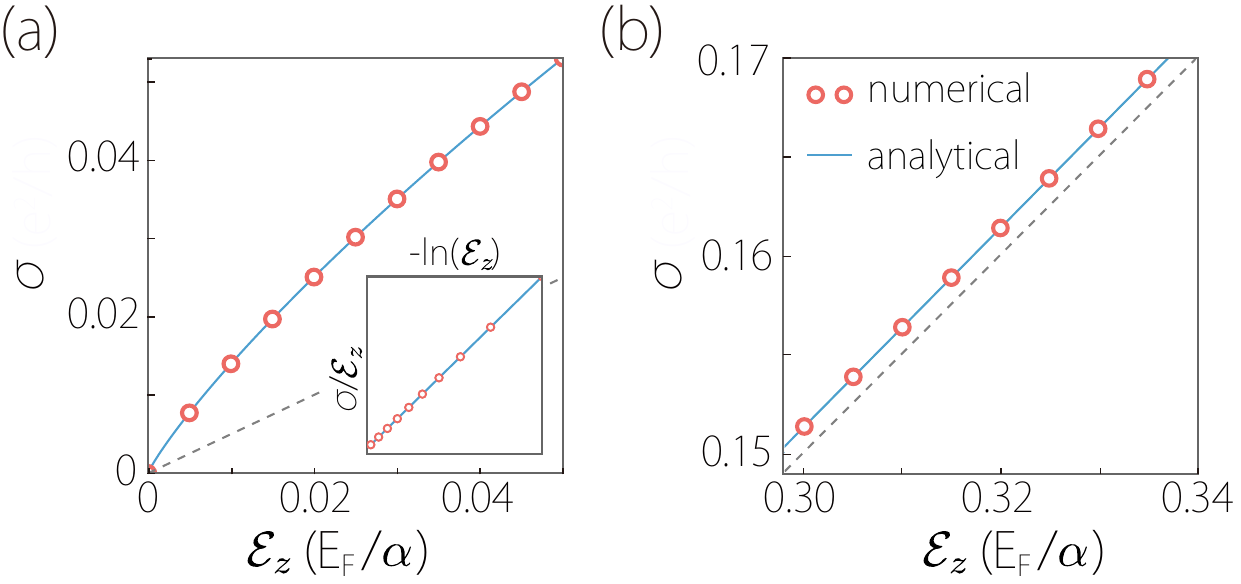}
	\caption{EHE response $\sigma_{xy}^{(T)}$ at finite temperature. 
		(a) Logarithmically corrected scaling in the weak-$\Ez$ limit. 
		The inset plots $\sigma_{xy}^{(T)}/\Ez$ as a function of $-\ln(\Ez)$, where the logarithmically corrected scaling is manifested as a linear relation with a finite slope. 
		(b) Linear behavior of $\sigma_{xy}^{(T)}$ in the low-temperature limit. 
		In both panels, the red open circles and blue solid curves denote the finite-temperature numerical and analytic results, respectively, while the gray dashed curves show the corresponding zero-temperature results.
		We set ${\cal C}=1$, $E_F=10~\mathrm{meV}$, and $\alpha=1~\text{\AA}$, with $\kb T=5~\mathrm{meV}$ for (a) and $\kb T=0.5~\mathrm{meV}$ for (b).
	}
	\label{Fig.3}
\end{figure}

We have a few remarks here. First, Eq.~(\ref{eq:sigma_log}) shows that $\sigma_{xy}^{(T)}$ still vanishes at $\Ez=0$, as $\ln (1/|\Ez|)\Ez=0$ for $\Ez=0$, consistent with the underneath physics: the Hall current is induced by  $\Ez$. 
Second, Eq.~(\ref{eq:sigma_log}) is not the common linear but an unconventional logarithmically corrected scaling  of $\Ez$, further demonstrating the highly nontrivial feature  of the EHE in magnetic WPs. This unconventional  scaling is also confirmed by our numerical calculations, as shown in Fig.~\ref{Fig.3}(a).
Third, unlike the $|E_F|^{-1}$ scaling for $T=0$ , which requires a fine tuning of the Fermi energy to enhance the Hall signal, the logarithmic divergence at finite temperature  emerges for  any $E_F$. This means that the quantum critical signatures of EHE are not  smeared out but strengthened by thermal broadening, significantly facilitating the experimental detection and also the potential application of the EHE.
Fourth, the  prefactor of the scaling is still determined  by the winding number ${\cal C}$ and is not sensitive to the material details. 

\textit{\textcolor{blue}{Detecting weak electric field.}}
Beside its fundamental significance, the scaling-enhanced EHE of WPs has a direct implication for weak electric-field detection.
Hall-effect sensors are the most widely used magnetic-field sensors in current sensing technologies~\cite{Ripka2010,Khan2021}, because the ordinary Hall effect directly converts a magnetic field into a transverse electrical voltage, without relying on auxiliary transduction mechanisms. 
Analogously, the EHE provides an electric-field counterpart: an out-of-plane electric field is directly converted into a Hall response, as illustrated in Fig. \ref{Fig.1}(b).
More importantly, in 2D Weyl semimetals this conversion can be strongly amplified by the unconventional scaling of the EHE.
As the Fermi level approaches the Weyl point, the divergent electric-field susceptibility $\chi$ ensures that even a weak electric field can generate a measurable Hall signal.
Therefore, the mature Hall readout scheme, together with the scaling-enhanced strong EHE response, may offer a new route toward compact all-electric detection of weak electric field.

To assess this possibility, we provide a simple estimation for the detection of $\Ez$ based on Eq.~(\ref{eq:sigma_full}). 
Taking $\alpha=5~\text{\AA}$, ${\cal C}=1$ for a linear WP, $E_F=2\mathrm{meV}$, and $T=10\mathrm{K}$, we find that a detectable Hall conductivity of $\sigma_{xy}=10^{-5}e^2/h$ corresponds to an electric field as weak as $\Ez=0.15\mathrm{V/cm}$. 
This value is comparable to those achieved by existing electric-field detection technologies~\cite{Liu2024EField}, suggesting that the EHE can serve as a feasible mechanism for electrically detecting weak electric fields.

\textit{\textcolor{blue}{Discussion.}}
In this work, we have established  the scaling of  EHE in  WPs. At zero temperature, the $\Ez$-induced Hall conductivity  linearly scales with $\Ez$ but features a divergent  $|E_F|^{-1}$ scaling, with the prefactor being determined by the global rather than local quantities of system. 
At finite  temperature, the scaling of  EHE evolves into an unconventional logarithmically corrected scaling  of $\Ez$.
This logarithmically corrected behavior, however, crosses over to the conventional linear scaling when the temperature is low and $E_F$ is finite.
In the limit of  low temperature $E_F-|\alpha \Ez| \gg k_BT$, we can expand $\sigma_{xy}^{(T)}$  in the series of $k_BT$ via the  standard Sommerfeld expansion~\cite{Ashcroft1976,XiaoYaoFangNiu2006},  
\begin{align}\label{SigmaLinear}
	&\sigma_{xy}^{(T)}(E_F)
	=	\int_{-\infty}^{\infty}d\varepsilon \sigma_{xy}^{(0)}(\varepsilon)  {\cal B}(\varepsilon-E_F,T)  \nonumber\\
&\simeq \frac{\cal C}{2}	\times \frac{\alpha\Ez}{|E_F|}\left[1+\frac{\pi^2}{3} \left(\frac{\kb T}{E_F}\right)^2\right] + O(k_B^4T^4),
\end{align}
which linearly scales with $\Ez$, and in the leading order reduces to the result of zero temperature.
From Eq.~(\ref{SigmaLinear}), we can more clearly observe that the temperature effect has an enhancing rather than a suppressing influence on EHE, as manifested in the  temperature-enhanced coefficient for $\Ez$ [see Fig.~\ref{Fig.3}(b)].

The  zero- and finite-temperature scalings of EHE provide a practical method to identify  2D Weyl semimetals. 
For a trivial  2D  system, its  Hall conductivity generally linearly increase or decreases with  $\Ez$, whereas a 2D WP gives the unconventional  logarithmically corrected scaling. 
More importantly, this  scaling does not  require the Fermi energy  to be  tuned to around the WP.
By further obtaining the EHE signal as  a function of  Fermi energy, we can locate the WP in energy.

In above discussions, we only focus on the EHE of a single WP. For realistic materials, the number of WPs can be more than one, and the WPs may exhibit opposite winding number.
However, as long as the magnetic Weyl semimetals belonging to one of the target groups listed in Table 	\ref{TabMLG}, a net EHE  should persist.

\begin{table}[t]
	\caption{Magnetic layer groups (MLGs) hosting both symmetry-stabilized WP and intrinsic EHE.
LWP, QWP and CWP denote the WPs with  linear, quadratic, and cubic dispersion, respectively. $\bm{k}$ denotes the high-symmetry points and lines.}
	\label{TabMLG}
	\begin{ruledtabular}
		\begin{tabular}{@{}lll@{\hspace{1.3em}}lll@{}}
			MLG & $\bm{k}$ & Type & MLG & $\bm{k}$ & Type \\
			\hline
			5.3.19 & $A, B$ & LWP & 36.3.223 & $S$ & LWP \\
			8.1.34 & $\Lambda, V$ & LWP & & $\Delta, F$ & LWP \\
			9.1.41 & $\Lambda, V$ & LWP & 50.3.360 & $\Gamma, M$ & LWP \\
			10.1.45 & $\Lambda$ & LWP & 53.1.374 & $\Gamma, M, X$ & LWP \\
			19.1.104 & $\Gamma, X, Y, S$ & LWP & 54.1.381 & $\Gamma, M$ & LWP \\
			& $\Delta, D, \Sigma, C$ & LWP & & $\Delta, \Sigma, Y$ & LWP \\
			20.1.111 & $\Gamma, Y$ & LWP & 57.4.401 & $\Gamma, M, X$ & LWP \\
			& $\Delta, B, \Lambda$ & LWP & & $\Delta, Y$ & LWP \\
			21.1.118 & $\Gamma, S$ & LWP & 58.4.408 & $\Gamma, M$ & LWP \\
			& $\Delta, D, \Sigma, C$ & LWP & & $\Delta, Y$ & LWP \\
			22.1.122 & $\Gamma, Y$ & LWP & 59.5.414 & $\Gamma, M$ & LWP \\
			& $\Sigma, \Delta, F, C$ & LWP & & $\Sigma$ & LWP \\
			27.3.156 & $\Lambda, H$ & LWP & 60.5.421 & $\Gamma$ & LWP \\
			28.3.169 & $U, Z$ & LWP & & $\Delta, \Sigma, Y$ & LWP \\
			& $\Lambda, G$ & LWP & 68.1.470 & $\Gamma, K$ & LWP \\
			29.3.176 & $\Lambda, H$ & LWP & & $T, T'$ & LWP \\
			30.3.183 & $\Lambda, H$ & LWP & 74.3.494 & $\Gamma$ & QWP \\
			31.3.190 & $Y, T$ & LWP & 76.1.500 & $\Gamma$ & CWP \\
			& $\Lambda, H$ & LWP & & $\Gamma, M, K$ & LWP \\
			32.3.199 & $Y, Z$ & LWP & & $T, T', \Sigma$ & LWP \\
			& $\Lambda, H$ & LWP & 78.5.514 & $\Gamma$ & QWP \\
			33.3.204 & $Y$ & LWP & & $\Sigma$ & LWP \\
			& $\Lambda, H$ & LWP & 79.4.518 & $\Gamma$ & QWP \\
			34.3.209 & $Y$ & LWP & & $K$ & LWP \\
			& $\Lambda, H$ & LWP & & $T, T'$ & LWP \\
			35.3.214 & $\Delta, F$ & LWP & & & \\
		\end{tabular}
	\end{ruledtabular}
\end{table}

\bibliography{ref}

\end{document}


\title{Supplemental Material for ``Unconventional Scaling of Electric Hall Effect in Magnetic Weyl Semimetals''}

\author{Chaoxi Cui}
\affiliation{Centre for Quantum Physics, Key Laboratory of Advanced Optoelectronic Quantum Architecture and Measurement (MOE), School of Physics, Beijing Institute of Technology, Beijing 100081, China}
\affiliation{Beijing Key Lab of Nanophotonics \& Ultrafine Optoelectronic Systems, School of Physics, Beijing Institute of Technology, Beijing 100081, China}

\author{Yilin Han}
\affiliation{Centre for Quantum Physics, Key Laboratory of Advanced Optoelectronic Quantum Architecture and Measurement (MOE), School of Physics, Beijing Institute of Technology, Beijing 100081, China}
\affiliation{Beijing Key Lab of Nanophotonics \& Ultrafine Optoelectronic Systems, School of Physics, Beijing Institute of Technology, Beijing 100081, China}

\author{Run-Wu Zhang}
\affiliation{Centre for Quantum Physics, Key Laboratory of Advanced Optoelectronic Quantum Architecture and Measurement (MOE), School of Physics, Beijing Institute of Technology, Beijing 100081, China}
\affiliation{Beijing Key Lab of Nanophotonics \& Ultrafine Optoelectronic Systems, School of Physics, Beijing Institute of Technology, Beijing 100081, China}

\author{Zhi-Ming Yu}
\email{zhiming\_yu@bit.edu.cn}
\affiliation{Centre for Quantum Physics, Key Laboratory of Advanced Optoelectronic Quantum Architecture and Measurement (MOE), School of Physics, Beijing Institute of Technology, Beijing 100081, China}
\affiliation{Beijing Key Lab of Nanophotonics \& Ultrafine Optoelectronic Systems, School of Physics, Beijing Institute of Technology, Beijing 100081, China}

\author{Yugui Yao}
\affiliation{Centre for Quantum Physics, Key Laboratory of Advanced Optoelectronic Quantum Architecture and Measurement (MOE), School of Physics, Beijing Institute of Technology, Beijing 100081, China}
\affiliation{Beijing Key Lab of Nanophotonics \& Ultrafine Optoelectronic Systems, School of Physics, Beijing Institute of Technology, Beijing 100081, China}

\maketitle

\tableofcontents

\section{Mapping from \texorpdfstring{$k$}{k} space to \texorpdfstring{$d$}{d} space}
\label{secS_mapping}

\begin{figure}[t]
\centering
\includegraphics[width=0.6\columnwidth]{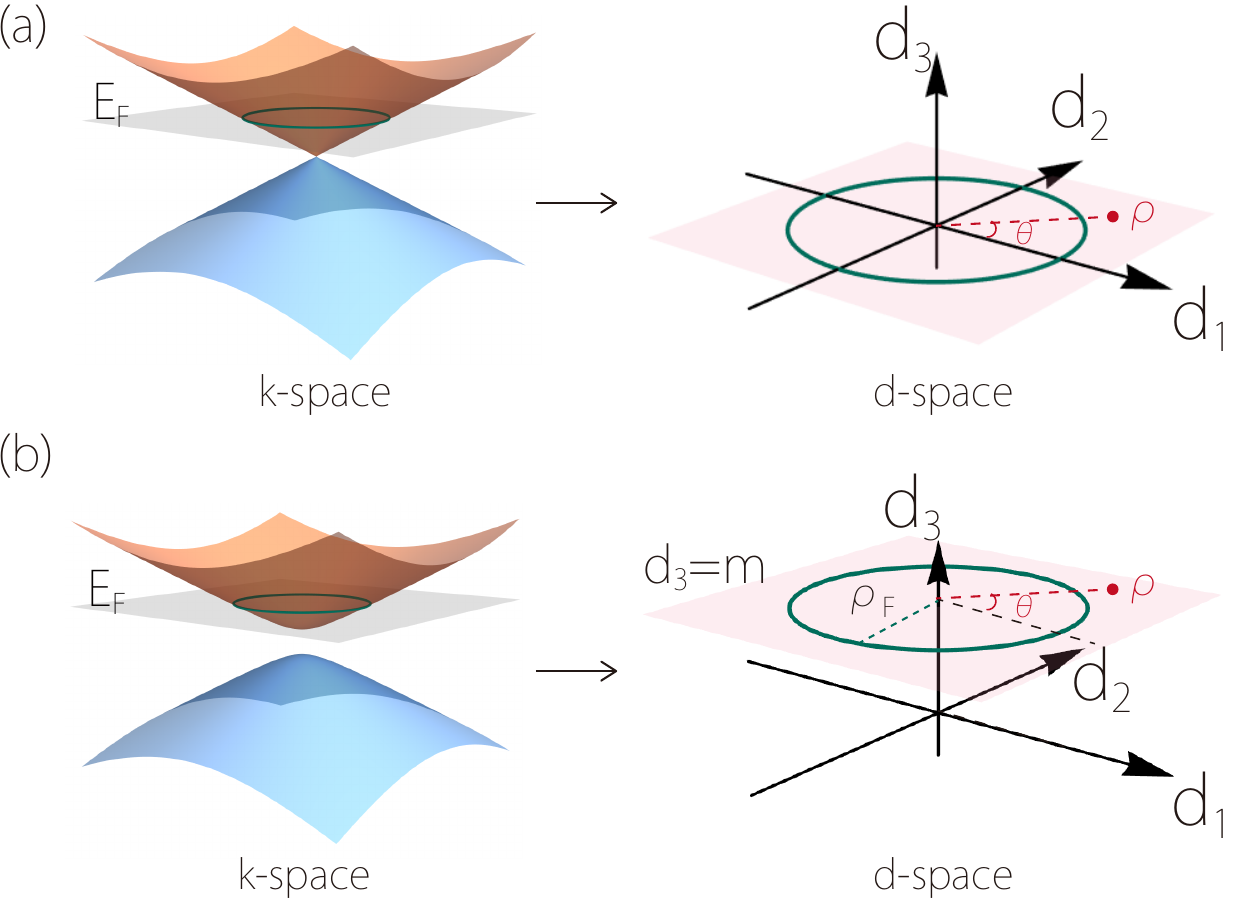}
\caption{Schematic of the mapping defined by (a) Eq.~(\ref{w/oEzmapping}) and (b) Eq.~(\ref{eqS_d1d2_from_F}).}
\label{fig1}
\end{figure}

In this section, we formulate the mapping from momentum space ($k$ space) to the space of Pauli-matrix coefficients ($d$ space)~\cite{Asboth2016ShortCourse}, which underlies the derivations in the following sections. A generic two-band Hamiltonian can be written as
\begin{equation}
H(\bm k)=d_1(\bm k)\tau_1+d_2(\bm k)\tau_2+d_3(\bm k)\tau_3,
\qquad
\bm d=(d_1,d_2,d_3).
\label{eqS_general_two_band}
\end{equation}
A scalar term $d_0\tau_0$ is omitted because it only shifts the two eigenvalues equally and affects neither the eigenstates nor the Berry curvature. 
The coefficients $\bm d(\bm k)$ define the relevant map from $k$ space to $d$ space
\begin{equation}
	\bm k\longmapsto\bigl(d_1(\bm k),d_2(\bm k),d_3(\bm k)\bigr).
	\label{kdmapping}
\end{equation}
For the Weyl points (WPs) considered in the main text, the symmetry-protected gapless Hamiltonian has $d_3=0$ at $\Ez=0$ and reads~\cite{Zhang2023Encyclopedia}
\begin{equation}
H_0(\bm k)=
F(\bm k)\tau_+ + F^*(\bm k)\tau_-,
\qquad
F(\bm k)=c_1 k_-^\nu+c_2 k_+^\nu,
\label{eqS_H_aniso_main}
\end{equation}
where $k_\pm=k_x\pm i k_y$, $\tau_\pm=(\tau_1\pm i\tau_2)/2$, and $\nu=1,2,3$ for linear, quadratic, and cubic WPs, respectively. Its winding number is ${\cal C}=\nu\eta=\nu\sgn(|c_1|^2-|c_2|^2)$. Applying the mapping in Eq.~(\ref{kdmapping}), we obtain
\begin{equation}
d_1(\bm k)=\mathrm{Re}\,F(\bm k),\qquad d_2(\bm k)=-\mathrm{Im}\,F(\bm k),\qquad d_3(\bm k)=0,
\label{w/oEzmapping}
\end{equation}
which maps the two-dimensional $k$ space into the $d_3=0$ plane in $d$ space. This mapping is illustrated in Fig.~\ref{fig1}(a). 


The out-of-plane electric field introduces the  mass term
\begin{equation}
H_{\mathcal E}=\Ez\hat z=m\tau_3,
\qquad
m=\alpha\Ez,
\label{eqS_mass_definition}
\end{equation}
where $\hat z=\alpha\tau_3$ is  the layer-coordinate operator  in the two-band subspace. The field-dependent Hamiltonian then becomes
\begin{equation}
H(\bm k,\Ez)=d_1(\bm k)\tau_1+d_2(\bm k)\tau_2+m\tau_3.
\label{eqS_H_massive}
\end{equation}
The electric field lifts the planar map in Eq.~(\ref{w/oEzmapping}) to a map from  $k$ space to the constant-$d_3$ plane in  $d$ space, as shown in Fig.~\ref{fig1}(b):
\begin{equation}
	d_1(\bm k)=\mathrm{Re}\,F(\bm k),\qquad d_2(\bm k)=-\mathrm{Im}\,F(\bm k),\qquad d_3(\bm k)=m.
	\label{eqS_d1d2_from_F}
\end{equation}
We next relate the Berry flux in $k$ space to that in $d$ space. Consider a patch $S^k$ in $k$ space and denote its image by $S^d$. At fixed $d_3$, the chain rule gives
\begin{equation}
\begin{aligned}
|\partial_{k_x}u_n\rangle
&=\frac{\partial d_1}{\partial k_x}|\partial_{d_1}u_n\rangle
+\frac{\partial d_2}{\partial k_x}|\partial_{d_2}u_n\rangle,\\
|\partial_{k_y}u_n\rangle
&=\frac{\partial d_1}{\partial k_y}|\partial_{d_1}u_n\rangle
+\frac{\partial d_2}{\partial k_y}|\partial_{d_2}u_n\rangle.
\end{aligned}
\label{eqS_chain_rule}
\end{equation}
We define the scalar Berry curvatures normal to the $k_x$-$k_y$ and $d_1$-$d_2$ planes, respectively, as
\begin{equation}
\Omega^k_n
=-2\,\mathrm{Im}\langle\partial_{k_x}u_n|\partial_{k_y}u_n\rangle,
\qquad
\Omega^d_n
=-2\,\mathrm{Im}\langle\partial_{d_1}u_n|\partial_{d_2}u_n\rangle.
\label{eqS_Berry_components}
\end{equation}
Here, $\Omega_n^k$ is the scalar Berry curvature denoted by $\Omega_n$ in Eq.~(5) of the main text.
Substituting Eq.~(\ref{eqS_chain_rule}) into Eq.~(\ref{eqS_Berry_components}) yields
\begin{equation}
\Omega^k_n=J_{d/k}\,\Omega^d_n,
\qquad
J_{d/k}=\det
\begin{bmatrix}
\partial_{k_x}d_1 & \partial_{k_y}d_1\\
\partial_{k_x}d_2 & \partial_{k_y}d_2
\end{bmatrix}.
\label{eqS_Jacobian_relation}
\end{equation}
Because $dS^d=|J_{d/k}|\,dS^k$, the flux element satisfies
\begin{equation}
\Omega_n^k\,dS^k
=\sgn(J_{d/k})\,
\Omega_n^d\,dS^d.
\label{eqS_flux_element}
\end{equation}
Thus, the local stretching factor cancels between the Berry curvature and the area element, while $\sgn(J_{d/k})$ records whether the map preserves or reverses orientation. Because the mapping from $k$ space to $d$ space is a ${|\cal C|}$-to-one mapping, the total flux satisfies
\begin{equation}
\int_{S^k}\Omega_n^k\,dS^k
={\cal C}\int_{S^d}\Omega_n^d\,dS^d.
\label{eqS_flux_mapping_general}
\end{equation}

\section{Zero-temperature EHE coefficient of Weyl points}
\label{secS_chi_aniso}

We now derive the zero-temperature electric Hall effect (EHE) coefficient $\chi_{xy}^{(0)}$ of the $\Ez$-sensitive WPs. 
The EHE coefficient is defined as~\cite{Cui2025EHE}
\begin{align}
	\chi_{xy}=&\sum_n\int\frac{d^2k}{2\pi}
	\left(\partial_{\varepsilon_n} f \times
	P_{nn}\Omega_n
	+f \Lambda_n
	\right).
	\label{eq:chi}
\end{align}
The Hamiltonian in Eq.~(\ref{eqS_H_aniso_main}) has two eigenvalues,
\begin{equation}
\varepsilon_s=s\rho, \quad \rho(\bm k)=\sqrt{d_1^2(\bm k)+d_2^2(\bm k)},
\label{eqS_energy_s}
\end{equation}
where $s=+1$ and $s=-1$ label the conduction and valence bands, respectively. A convenient gauge choice for the corresponding eigenstates is
\begin{equation}
|u_s\rangle=\frac{1}{\sqrt2}
\begin{pmatrix}
1\\
s e^{i\theta}
\end{pmatrix}, \quad \cos\theta(\bm k)=\frac{d_1(\bm k)}{\rho(\bm k)},
\quad
\sin\theta(\bm k)=\frac{d_2(\bm k)}{\rho(\bm k)}.
\label{eqS_eigenstate_s}
\end{equation}
For the $\Ez$-sensitive WPs considered in the main text, the effective layer-polarization matrix elements are $P_{ss'}=\langle u_s|\hat z|u_{s'}\rangle
=\alpha\langle u_s|\tau_3|u_{s'}\rangle$. Therefore,
\begin{equation}
P_{ss}=0,
\qquad
P_{s,-s}=\alpha,
\label{eqS_P_matrix}
\end{equation}
which implies that the first term in Eq.~(\ref{eq:chi}) vanishes. The velocity matrix elements $v_a^{ss'}=\langle u_s|\partial_{k_a}H_0|u_{s'}\rangle$ are
\begin{equation}
v_a^{ss}=s\partial_{k_a}\rho,
\qquad
v_a^{s,-s}=is\rho\,\partial_{k_a}\theta,
\qquad
v_a^{-s,s}=-is\rho\,\partial_{k_a}\theta.
\label{eqS_velocity_matrix}
\end{equation}
For a two-band model, the general expression for $\Lambda_s$ becomes~\cite{Cui2025EHE}
\begin{equation}
\begin{aligned}
\Lambda_s
=
2\mathrm{Im}
\frac{
2\delta P_{s\bar s}v_x^{s\bar s}v_y^{\bar s s}
+P_{\bar s s}\delta V_y^{s\bar s}v_x^{s\bar s}
+P_{s\bar s}\delta V_x^{s\bar s}v_y^{\bar s s}
}{
(\delta\varepsilon_{s\bar s})^3
},
\end{aligned}
\label{eqS_Lambda_two_band}
\end{equation}
where $\bar s=-s$ and
\begin{equation}
\delta\varepsilon_{s\bar s}\equiv\varepsilon_s-\varepsilon_{\bar s}=2s\rho,
\qquad
\delta P_{s\bar s}\equiv P_{ss}-P_{\bar s\bar s}=0,
\qquad
\delta V_a^{s\bar s}\equiv v_a^{ss}-v_a^{\bar s\bar s}=2s\partial_{k_a}\rho.
\label{eqS_two_band_differences}
\end{equation}
Using Eqs.~(\ref{eqS_P_matrix}) and (\ref{eqS_velocity_matrix}), Eq.~(\ref{eqS_Lambda_two_band}) gives
\begin{equation}
\begin{aligned}
\Lambda_s(\bm k)
&=-\frac{s\alpha}{2\rho^2}
\left(
\partial_{k_x}\rho\,\partial_{k_y}\theta
-\partial_{k_y}\rho\,\partial_{k_x}\theta
\right),
\end{aligned}
\label{eqS_Lambda_rho_theta}
\end{equation}
where $\rho
\left(
\partial_{k_x}\rho\,\partial_{k_y}\theta
-\partial_{k_y}\rho\,\partial_{k_x}\theta
\right)$ is precisely the Jacobian determinant $J_{d/k}$ in Eq.~(\ref{eqS_Jacobian_relation}). Therefore, we obtain the compact expression
\begin{equation}
\Lambda_s(\bm k)=-\frac{s\alpha}{2}\frac{J_{d/k}(\bm k)}{\rho^3(\bm k)}.
\label{eqS_Lambda_J}
\end{equation}

For $E_F>0$, the valence band is fully occupied, whereas the conduction band is occupied only for $\rho<E_F$. Because $\Lambda_+=-\Lambda_-$, the two-band contributions cancel in this region, leaving
\begin{equation}
	\chi_{xy}^{(0)}(E_F>0)
	=\frac{\alpha}{2}
	\int_{\rho>E_F}\frac{d^2k}{2\pi}\frac{J_{d/k}}{\rho^3}.
	\label{eqS_chi_integral_aniso}
\end{equation}
The remaining integral can be evaluated directly using the $k$-to-$d$ mapping established in Sec.~\ref{secS_mapping}. The area elements in $k$ space and $d$ space are related by
\begin{equation}
	dS^d\equiv dd_1\,dd_2
	=|J_{d/k}|\,d^2k.
	\label{eqS_area_mapping_chi}
\end{equation}
The factor $|J_{d/k}|$ in the area element cancels the magnitude of the Jacobian $J_{d/k}$ in the integrand, leaving only its sign. Accounting for both the orientation and the $|{\cal C}|$-to-one multiplicity of the mapping, we obtain
\begin{align}
	\int_{\rho>E_F}d^2k\,\frac{J_{d/k}}{\rho^3}
	={\cal C}\int_{\rho>E_F}\frac{dS^d}{\rho^3}
	={\cal C}\int_0^{2\pi}d\theta
	\int_{E_F}^{\infty}\frac{\rho\,d\rho}{\rho^3}
	=\frac{2\pi{\cal C}}{E_F}.
	\label{eqS_chi_k_to_d}
\end{align}
Substituting this into Eq.~(\ref{eqS_chi_integral_aniso}) yields
\begin{equation}
	\chi_{xy}^{(0)}(E_F>0)
	=\frac{\alpha}{2}\frac{{\cal C}}{E_F}.
	\label{eqS_chi_positive}
\end{equation}
For $E_F<0$, the conduction band is empty, and only the valence-band states with $\rho>|E_F|$ are occupied. A straightforward calculation gives
\begin{equation}
	\chi_{xy}^{(0)}(E_F<0)
	=-\frac{\alpha}{2}\frac{{\cal C}}{E_F}.
		\label{eqS_chi_negitive}
\end{equation}
Combining Eqs.~(\ref{eqS_chi_positive}) and (\ref{eqS_chi_negitive}), we obtain Eq.~(9) of the main text:
\begin{equation}
	\chi_{xy}^{(0)}(E_F)
	=\frac{\alpha}{2}\frac{{\cal C}}{|E_F|}
	=\frac{\alpha}{2}\frac{\nu\eta}{|E_F|}.
\end{equation}

\section{Zero-temperature Hall conductivity of Weyl points}
\label{secS_sigma_aniso}
We next derive the zero-temperature Hall conductivity of $\Ez$-sensitive WPs as a function of the out-of-plane electric field $\Ez$. For the massive Hamiltonian, $d_3=m=\alpha\Ez$. On the constant-$d_3$ plane, we introduce polar coordinates
\begin{equation}
d_1=\rho\cos\theta,
\qquad
d_2=\rho\sin\theta,
\qquad
d=\sqrt{\rho^2+m^2}, 
\label{eqS_d_polar_sigma}
\end{equation}
as shown in Fig.~\ref{fig1}(b). The scalar Berry curvatures of the conduction and valence bands, normal to the constant-$d_3$ plane, are~\cite{XiaoChangNiu2010}
\begin{equation}
\Omega_c^d=-\frac{d_3}{2d^3}
=-\frac{m}{2(\rho^2+m^2)^{3/2}},
\qquad
\Omega_v^d=\frac{d_3}{2d^3}
=\frac{m}{2(\rho^2+m^2)^{3/2}}.
\label{eqS_Berry_d}
\end{equation}
The valence-band flux through the entire constant-$d_3$ plane is
\begin{equation}
\int_{d_3=m}\Omega_v^d\,dS^d
=\int_0^{2\pi}d\theta\int_0^\infty \rho d\rho\,
\frac{m}{2(\rho^2+m^2)^{3/2}}
=\pi\sgn(m).
\label{eqS_full_flux_sigma}
\end{equation}
Using Eq.~(\ref{eqS_flux_mapping_general}), the Hall conductivity can be written in terms of the occupied-state Berry curvature as~\cite{XiaoChangNiu2010}
\begin{equation}
\sigma_{xy}^{(0)}
=\frac{1}{2\pi}\sum_n\int_{S_{n,\mathrm{occ}}^k}
\Omega_n^k\,dS^k
=\frac{{\cal C}}{2\pi}\sum_n\int_{S_{n,\mathrm{occ}}^d}
\Omega_n^d\,dS^d,
\label{eqS_sigma_flux_general}
\end{equation}
where $S_{n,\mathrm{occ}}^k$ and $S_{n,\mathrm{occ}}^d$ denote the occupied region of band $n$ in $k$ space and its image in $d$ space, respectively. If the Fermi energy lies inside the field-induced gap, $|E_F|<|m|$, the occupied flux is the full valence-band flux. Equations~(\ref{eqS_full_flux_sigma}) and (\ref{eqS_sigma_flux_general}) then give
\begin{equation}
\sigma_{xy}^{(0)}(E_F,\Ez)
=\frac{{\cal C}}{2}\sgn(m),
\qquad |E_F|<|m|.
\label{eqS_sigma_gap}
\end{equation}
When $|E_F|>|m|$, the Fermi contour in $d$ space has radius
\begin{equation}
\rho_F=\sqrt{E_F^2-m^2}.
\label{eqS_varrho_F}
\end{equation}
For $E_F>0$, the valence band is fully occupied, and the conduction band is occupied within $\rho\leq\rho_F$. The additional conduction-band flux is
\begin{equation}
\begin{aligned}
\int_0^{2\pi}d\theta\int_0^{\rho_F}\rho d\rho\,
\Omega^d_c
=-\int_0^{2\pi}d\theta\int_0^{\rho_F}\rho d\rho\,
\frac{m}{2(\rho^2+m^2)^{3/2}}=-\pi\left[\sgn(m)-\frac{m}{E_F}\right].
\end{aligned}
\label{eqS_conduction_flux_sigma}
\end{equation}
Adding the contribution of the fully occupied valence band gives the Hall conductivity for $E_F>0$:
\begin{equation}
	\sigma_{xy}^{(0)}(E_F,\Ez)
	=\frac{{\cal C}}{2}\frac{m}{E_F},
	\label{sigma_positive}
\end{equation}
For $E_F<0$, the unoccupied valence-band states contribute a flux opposite to that in Eq.~(\ref{eqS_conduction_flux_sigma}).
Subtracting this flux from that of the fully occupied valence band gives the same result as Eq.~(\ref{sigma_positive}). Therefore, when $|m|<|E_F|$, the Hall conductivity has the compact form
\begin{equation}
\sigma_{xy}^{(0)}(E_F,\Ez)
=\frac{{\cal C}}{2}\frac{m}{|E_F|},
\qquad |m|<|E_F|.
\label{eqS_sigma_metal}
\end{equation}
Combining the results of two regimes, and substituting $m=\alpha\Ez$, we obtain
\begin{equation}
\sigma_{xy}^{(0)}(E_F,\Ez)
=\frac{{\cal C}}{2}
\begin{cases}
\dfrac{\alpha\Ez}{|E_F|}, & |\alpha\Ez|\le|E_F|,\\[7pt]
\sgn(\alpha\Ez), & |\alpha\Ez|>|E_F|.
\end{cases}
\label{eqS_sigma_zeroT_piecewise}
\end{equation}

\section{Symmetry-allowed electric-field coupling of EHE-compatible WPs}
\label{secS_model_symmetry}

As discussed in the main text, the EHE-compatible WPs are obtained by intersecting the magnetic layer groups (MLGs) that allow the intrinsic EHE~\cite{Cui2025EHE} with those that support symmetry-protected two-dimensional magnetic Weyl points~\cite{Zhang2023Encyclopedia}. In this section, we determine how an out-of-plane electric field couples to these WPs. In the two-band basis, the general coupling takes the form
\begin{equation}
H_{\mathcal E}=\Ez\mathcal Z,
\qquad
\mathcal Z=z_0\tau_0+z_1\tau_1+z_2\tau_2+z_3\tau_3.
\label{eqS_HE}
\end{equation}
Here $\mathcal Z$ is the matrix representation of the layer-coordinate operator $\hat z$, $\tau_i$ are Pauli matrices, and the real coefficients $z_i$ depend on material details.
As established for intrinsic EHE~\cite{Cui2025EHE}, the relevant symmetry operations can only belong to
\begin{equation}
\mathcal S=\{E,M_{\parallel}\mathcal T,C_{n,z}\},
\qquad
M_z\mathcal T\mathcal S
=\{M_z\mathcal T,C_{2,\parallel},S_{n,z}\mathcal T\}.
\label{eqS_symmetry_sets}
\end{equation}
Here $E$ is the identity operation; $M_z$ and $M_{\parallel}$ are the horizontal and vertical mirrors, respectively; $C_{n,z}$ is an $n$-fold rotation about the out-of-plane axis, with $n=2,3,4,$ or $6$; $C_{2,\parallel}$ is an in-plane twofold rotation; $S_{n,z}$ is an improper rotation; and $\mathcal T$ is time reversal. The operations in $\mathcal S$ preserve the layer coordinate, whereas those in $M_z\mathcal T\mathcal S$ reverse it. Therefore, $\mathcal Z$ satisfies
\begin{equation}
\begin{aligned}
\mathcal O\mathcal Z\mathcal O^{-1}&=\mathcal Z,
&\qquad \mathcal O&\in \mathcal S,\\
\mathcal O\mathcal Z\mathcal O^{-1}&=-\mathcal Z,
&\qquad \mathcal O&\in M_z\mathcal T\mathcal S.
\end{aligned}
\label{eqS_Z_constraints}
\end{equation}

Applying Eq.~(\ref{eqS_Z_constraints}) to all $\Ez$-sensitive WPs listed in Table I of the main text gives two general results:
\begin{itemize}
	\item The $\tau_0$ term is excluded by the transformation rule under the layer-reversing operations in $M_z\mathcal T\mathcal S$.
	\item The $\tau_3$ term is symmetry allowed for every WP, whereas the $\tau_1$ and $\tau_2$ terms are forbidden in some MLGs.
\end{itemize}
The absence of the $\tau_0$ term means that the electric field does not shift the energy of the WP. The components proportional to $\tau_1$ and $\tau_2$ do not open a gap; instead, they shift the WP in the $d_1$-$d_2$ plane and leave the universal EHE scaling unchanged. The Hall-active, gap-opening component is therefore $z_3\tau_3$, which is the term retained in the main text.

We illustrate this procedure using the WP described by the irreducible representation $S_5$ at the $M$ point of MLG 58.4.408. Its low-energy Hamiltonian is~\cite{Zhang2023Encyclopedia}
\begin{equation}
H^{M}_{58.4.408}(\bm q)
=v(-q_x\tau_1+q_y\tau_2),
\label{eqS_H_584408}
\end{equation}
where $\bm q=(q_x,q_y)$ is measured from the $M$ point and $v$ is a real dispersion coefficient. The generators of the little group are $C_{2z}, C_{2x}$ and $S_{4z}\mathcal T$, whose matrix representations are~\cite{Zhang2023Encyclopedia}
\begin{equation}
C_{2z}=-i\tau_3,
\qquad
C_{2x}=-\tau_1,
\qquad
S_{4z}\mathcal T=
\begin{pmatrix}
0&1\\
-i&0
\end{pmatrix}\mathcal K,
\label{eqS_symmetry_584408}
\end{equation}
where $\mathcal K$ denotes complex conjugation.
For this WP, $C_{2z}\in\mathcal S$, whereas $C_{2x}$ and $S_{4z}\mathcal T$ belong to $M_z\mathcal T\mathcal S$. Equation~(\ref{eqS_Z_constraints}) therefore becomes
\begin{equation}
\begin{aligned}
C_{2z}\mathcal ZC_{2z}^{-1}&=\mathcal Z,\\
C_{2x}\mathcal ZC_{2x}^{-1}&=-\mathcal Z,\\
(S_{4z}\mathcal T)\mathcal Z(S_{4z}\mathcal T)^{-1}&=-\mathcal Z.
\end{aligned}
\label{eqS_Z_constraints_584408}
\end{equation}
Substituting Eq.~(\ref{eqS_HE}) into these three constraints gives $z_0=z_1=z_2=0$ and leaves
\begin{equation}
\mathcal Z^{M}_{58.4.408}=z_3\tau_3,
\qquad
H_{\mathcal E,58.4.408}^{M}
=\Ez\mathcal Z^{M}_{58.4.408}
=\alpha\Ez\tau_3
=m\tau_3.
\label{eqS_Z_HE_584408}
\end{equation}

\bibliography{refS}